\documentclass{elsart}

\usepackage{graphicx}

\usepackage{amssymb}

\begin{document}

\begin{frontmatter}

\title{Photon yields from nitrogen gas and dry air excited by electrons}

\author[FUKUI-SCE]{M. Nagano},
\author[FUKUI-ACE]{K. Kobayakawa},
\author[RIKEN]{N. Sakaki} and
\author[FUKUI-APC]{K. Ando}

\address[FUKUI-SCE]{Department of Space Communication Engineering, Fukui
   University of Technology, Fukui, 910-8505 Japan}
\address[FUKUI-ACE]{Department of Architecture and Civil Engineering, 
   Fukui University of Technology, Fukui, 910-8505 Japan}
\address[RIKEN]{RIKEN (The Institute of Physical and Chemical Research), 
   Wako, 351-0198 Japan}
\address[FUKUI-APC]{Department of Applied Physics and
   Chemistry, Fukui University of Technology, Fukui, 910-8505 Japan}

\begin{abstract}
In order to detect ultrahigh-energy cosmic rays (UHECR), atmospheric
fluorescence light from the trajectory of the extensive air shower may
be measured by mirror-photosensor systems.  In this type of experiment
the photon yield from electrons exciting air of various densities and
temperatures is most fundamental information for estimating the
primary energy of UHECR.  An experiment has been undertaken using a
\nuc{90}{Sr} $\beta$ source to study the pressure dependence of photon
yields, and the life times of the excited states, for radiation in
nitrogen and dry air.  The photon yield between 300 nm and 406 nm in
air excited by 0.85 MeV electrons is 3.73$\pm$0.15 ($\pm$14 \%
systematic) photons per meter at 1000 hPa and 20 $^{\circ}$C.  The air
density and temperature dependence is given for application to UHECR
observations.
\end{abstract}

\end{frontmatter}

\section{Introduction}
\label{sec:intro}
The fluorescence technique was first successfully used by the Fly's
Eye detector to explore cosmic rays in the ultra-high energy region
\cite{bal85}. From 1998 the HiRes detector, a successor to the Fly's
Eye, has been in operation with improved resolution in energy, arrival
directions and in meaurements of the longitudinal shower development
\cite{abu99}.  Fluorescence detectors will be constructed along with
the surface array at the Pierre Auger Observatory \cite{ces01} and one
of them is now in operation in Malargue, Argentina.  In these
experiments an atmosphere of quite low vapour pressure (i.e. dry) is used
as a vast scintillator.  In contrast, in the case of a satellite-based
telescope viewing downward (EUSO) \cite{sca01}, fluorescence light
from extensive air showers is observed mainly over the ocean where air
is not dry.

While the fluorescence efficiencies of electrons in the troposphere
are fundamentally important, there are only a few measurements which
can be applied to the UHECR experiments.  Bunner summarized in his
Ph.D thesis \cite{bun64} the fluorescence radiation from air and
showed that photons from the 2nd positive (2P) band of the nitrogen
molecule and the 1st negative (1N) band of the nitrogen ion were
significant.  Since there are many vibrational states, there are many
lines between 300 nm and 400 nm.  Photon yields listed in Bunner's
thesis are based on those excited by low energy electrons, deuterons
and alpha particles stopped in air (i.e. total absorption conditions)
and the errors of each experiment are about 30 \%.  It is important to
measure the photon yields excited by electrons in various bands under
thin target conditions. In order to apply them to air shower
observations, their density and temperature dependencies must be known
in each band.

Kakimoto et al. \cite{kak96} measured photon yields in thin targets
and showed that the yields were proportional to the particle energy
loss per unit length, $\d E/\d x$, between 1.4 MeV and 1000 MeV.  However,
their measurements were limited to three main wavelength bands with
central wavelengths of 337 nm, 357 nm and 391 nm.  The results of
Kakimoto et al. are about 20--40 \% larger than those summarized by
Bunner at these three wavelengths.  They also measured the photon
yields using the HiRes optical filter which is sensitive in a wide
band between 300 and 400 nm.  Using this wide band measurement,
Kakimoto et al. estimate the absolute values of the lines not
individually measured, and they get the values lower than those
determined by Bunner.

Given these previous measurements, it is clear that precise
measurements at wavelengths other than the three major lines above are
required in thin targets, since optical filters between 300 nm and
406 nm are used in most cosmic ray experiments.  We measured the photon
yields of electrons in dry air with various pressures to apply to the
atmosphere in a desert.  The photon yields were measured through
filters of central wavelengths of 314.7, 337.7, 356.3, 380.9, 391.9
and 400.9 nm and 10 nm bandwidth.  Yields in damp air are now under
measurement and will be reported in a subsequent paper.

Recently, the HiRes group reported measurements of the energy spectrum
of UHECRs \cite{abu01}.  There are significant discrepancies in the
energy spectrum at the highest energies as measured by two
experiments, HiRes and AGASA (Akeno Giant Air Shower Array)
\cite{tak98}.  Using the values in the present measurement, we discuss
their effect on the energy determination of UHECR's by the
fluorescence technique.

\section{Experiment}
\label{sec:exp}
\subsection{Experimental arrangement}
\label{subsec:exp_arrange}
\begin{figure}[thb]
\centerline{\includegraphics[height=10cm]{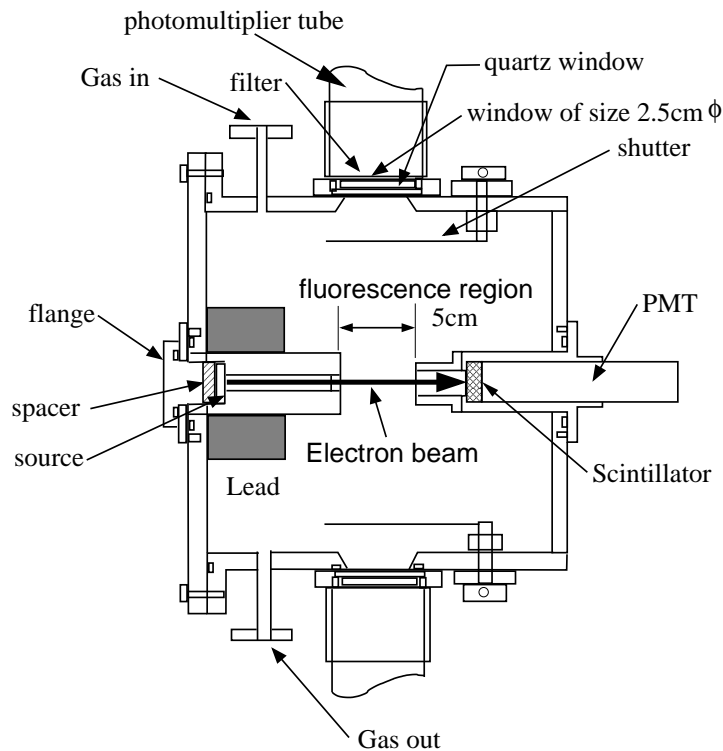}} 
\caption{Schematic drawing of the chamber (top view). Three PMTs
are mounted on two sides and the top of the chamber, and they view
fluorescence light through quartz windows.  Optical filters are
mounted between the PMTs and the windows. Electrons from
\nuc{90}{Y}$\rightarrow$\nuc{90}{Zr} are beamed and detected by a
scintillation counter.}
\label{chamber}
\end{figure}
\begin{figure}
\centerline{\includegraphics[height=8.3cm]{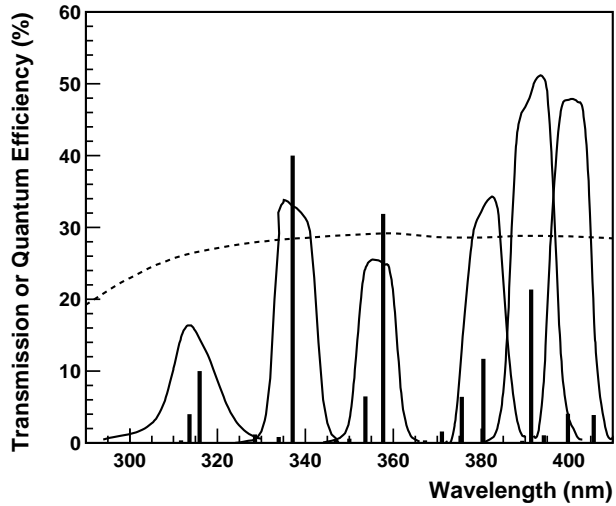}} 
\caption{Typical example of quantum efficiency of PMT used ( a dashed
curve) and transmission coefficients of interference filters used in
the present experiment.  The relative intensities of fluorescence
lines at 1000 hPa from the nitrogen molecule (solid lines) and ion
(dotted line) in air from Bunner \cite{bun64} are also shown neglecting their
spectral widths.}
\label{filter}
\end{figure}

We chose a photon counting and thin target technique to measure the
pressure dependence of photon yields (the number of photons produced
by electrons per meter of travel) from nitrogen and air excited by
electrons, following the method employed by Kakimoto et
al. \cite{kak96}.  The cubic chamber used is shown schematically in
Fig. \ref{chamber}. Three photomultiplier tubes of 2 inch diameter
(Hamamatsu photon counting H7195PX, named PM1, PM2 and PM3), which
were selected for low noise, were mounted on two sides and the top of
the chamber, and view the light through quartz windows.  
The window material of PMT is borosilicate glass and its typical quantum
efficiency is drawn by a dashed curve in Fig. \ref{filter}. 
The PMT high voltage and gain used in the measurements are about
1950V and 6$\times10^6$, respectively.
A central region (diameter 2.5 cm) of each PMT was used for photon counting.
Electrons with a maximum energy of 2.28 MeV from $\beta$ decay of
\nuc{90}{Sr}$\rightarrow$\nuc{90}{Y}$\rightarrow$\nuc{90}{Zr} (3.7 MBq)
 were beamed by a collimator.  The energy of each electron was measured 
by a scintillation counter (electron counter).  The coincidence between the
electron counter and any one of the photon counters was used to
generate a gate for the ADCs and a start signal for a TDC.  A beam
length of 5 cm was visible to the three photon counters and the
triggered rate of electrons was about 2.7$\times10^3$ s$^{-1}$ in
vacuum and 1.5$\times10^3$ s$^{-1}$ at 1000 hPa.  The electron energy
was reduced by about 0.17 MeV by a mylar window and the aluminum cover
of the source.

\begin{figure}[thb]
\centerline{\includegraphics[height=8.0cm]{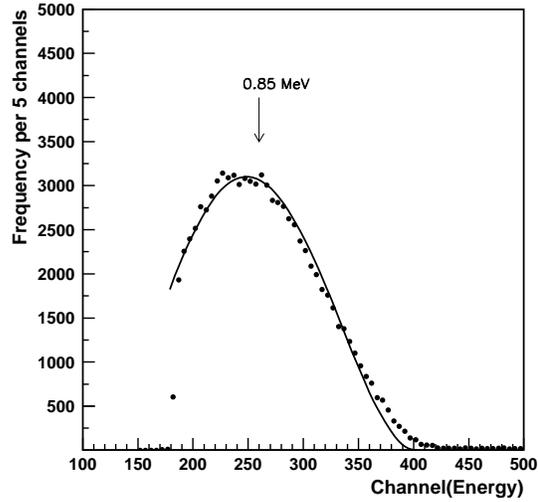}} 
\caption{Energy spectrum of the electrons. Dots represent the
observed spectrum and the solid curve is that expected from the
$\beta$ decay of \nuc{90}{Y}.  The threshold and average electron
energies are 0.3 MeV and 0.85 MeV respectively.}
\label{energy}
\end{figure}

The energy spectrum of the triggered electrons is shown in
Fig. \ref{energy}.  As seen there, the measured spectrum fits well to
the expected decay spectrum of $\beta$ rays depicted by the solid
curve.  The threshold and average electron energies are 0.3 and 0.85
MeV respectively.  The average energies below and above 0.85 MeV are
0.55 MeV and 1.2 MeV, respectively.  The difference in photon yields
between these two energy regions will also be presented.

In the present measurements, six narrow band interference filters with
central wavelengths of 314.7, 337.7, 356.3, 380.9, 391.9 and 400.9 nm
were used.  The bandwidth of each filter was about 10 nm.
 Hereafter we refer to 316, 337, 358, 380, 391
and 400 nm, respectively, to represent the main line in each filter band.
Fig. \ref{filter} shows the transmission coefficients of the adopted
filters as a function of wavelength (provided by the manufacturer,
Hi-Technology Inc).  Vertical lines show the relative photon yields of
fluorescence lines in air at 1000 hPa listed by Bunner \cite{bun64},
where solid and dotted lines correspond to the yield from 2P and 1N
bands, respectively.  There remain some lines which are not covered in
the present experiment, but their contribution to the total yield may
be less than 5 \%. 

In any one run under a particular set of conditions, the number of
incident electrons at the electron counter was registered.  If there
was a coincidence in 150 ns between the electron counter and any one
of three photon counters, the ADC values of the electron counter and
the corresponding photon counter were recorded.  The time difference
between the photon signal and the electron signal (which was delayed
by 180 ns) was also recorded by the TDC in units of 0.5 ns.

\begin{figure}
\centerline{\includegraphics[height=15.0cm]{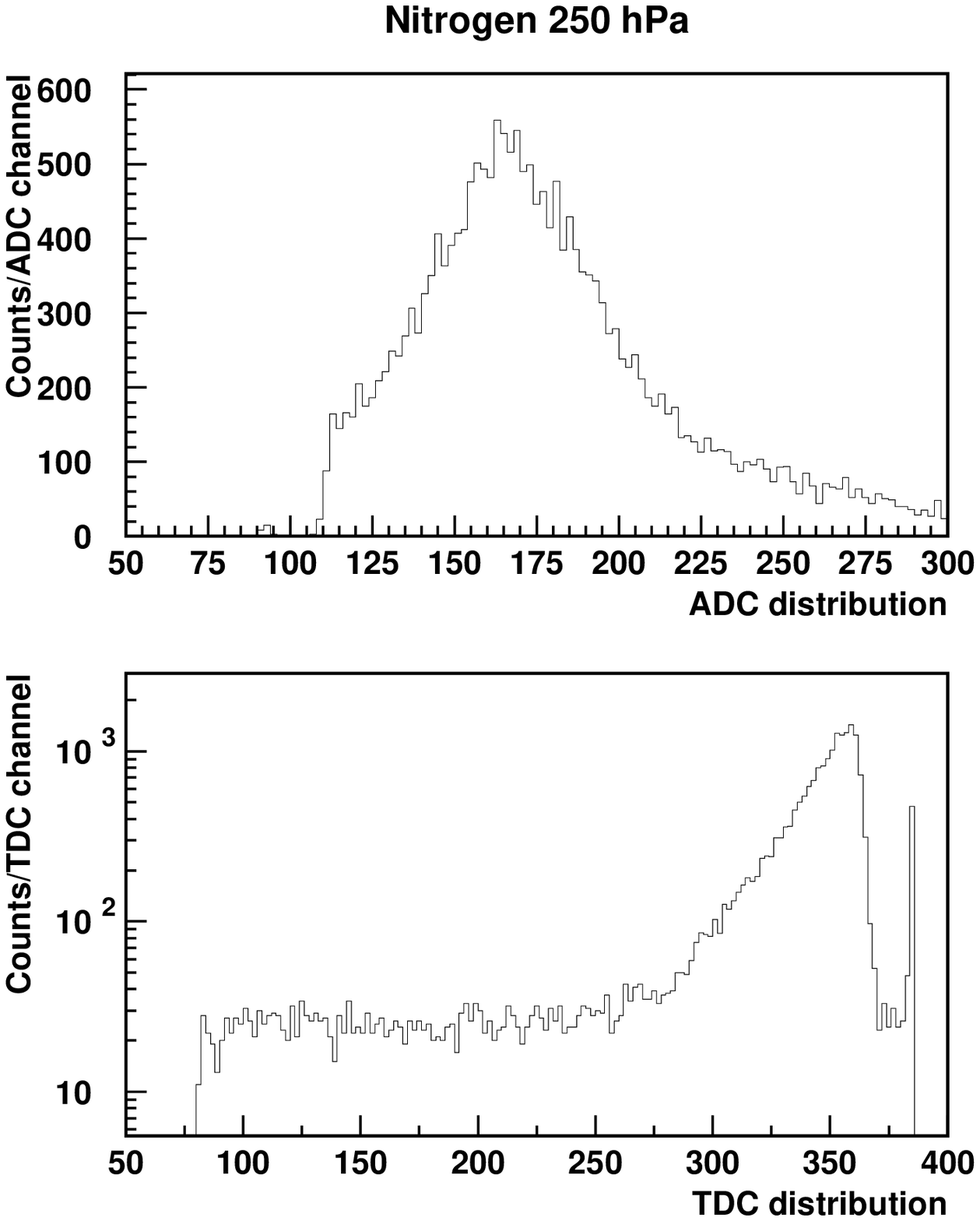}} 
\caption{An example of ADC and TDC distributions (for 250 hPa, N$_2$).
  The upper plot is a typical single photon pulse height
distribution and the lower plot shows the time difference between 
incident electrons and emitted photons. Each channel corresponds to 
0.25pC and 0.5 ns, respectively.
The TDC is started by the photon signal and is stopped by the electron
signal which is delayed by 180 ns.
}
\label{distribution}
\end{figure}

Examples of the ADC and TDC distributions are shown in
Fig. \ref{distribution}.  The coincident time of the electron counter
and the photon counter is shown as the 369th channel of the TDC.
Since the TDC is started by the photon signal and stopped by the
electron signal (delayed by 180 ns), the more the channel decreases
the more the photon signal is delayed.  The time resolution of the
system is measured by using coincident signals of cosmic ray muons
between PM2 at the top of the chamber and the electron counter, and is
found to be 1.26 ns.  Signal can clearly be separated
from the background in the TDC distribution.  Signals above the 369th
channel in the lower plot of Fig. \ref{distribution} are due to
background photon signals prior to the electron signal, which were
triggered due to the 20 ns pulse width of the discriminator output.

\subsection{Backgrounds and the difference from our past
measurements}
\label{subsec:backgrounds}
There is a background of single photon PMT noise, which is uniformly
distributed in the TDC distribution as shown in
Fig. \ref{distribution}.  With a shutter was placed in front of each
photon counter inside the chamber, some signals were observed even in
vacuum runs.
Their rates were (0.082 $\pm$ 0.06) min$^{-1}$ 
when a lead brick was not set around the source in
Fig. \ref{chamber}  and were reduced to $\frac{1}{8}$ with the
lead brick in the present measurement. 
This background  can,
 according to our calculations, be wholly attributed to
Bremsstrahlung photons from the aluminum foil (0.1 mm thickness) of the
source window.
 There are coincident muons measured by 
the photon counter at the top of the chamber (PM2) and
the electron counter.  This
rate is measured to be (0.133 $\pm$ 0.008) min$^{-1}$.  


The differences of past measurements reported
at the 27th ICRC (M1) \cite{nag01} and the present
ones (M2) are summarized as follows;

\begin{enumerate}
\item  The visible beam length was shortened from 10 cm to 5 cm,
  so that the average photon incident angle was reduced from 9.3$^{\circ}$
 to 5.2$^{\circ}$.  Though we corrected the interference filter transmission
 coefficients for these photon incident angles in M1, 
 the correction has not been applied to M2. The differences
 between both results are within 5\%, irrespective in
 any filter band.   
\item The background from Bremsstrahlung photons around the source region
in M1 has been reduced by about $\frac{1}{8}$
with the 5 cm thick lead brick around the source in M2.

\end{enumerate}

\subsection{Derivation of photon yields}
\label{subsec:photon}
The number of signal counts was obtained by subtracting backgrounds
determined from the background portion of the TDC
distribution. This background portion can be discriminated from the
signal portion as shown in Fig. \ref{distribution}.  The photon yield
per unit length per electron $\epsilon$, is determined as the number of
signal counts $N$ divided by the product of the following: the total
number of electrons $I$, the length of the fluorescence portion $a$,
the solid angle of the PMT $\Omega$, the quartz window transmission
$\eta$, the filter transmission $f$, and the quantum efficiency QE
and the collection efficiency CE of the PMT.

\begin{equation}
 \epsilon = \frac{N}{I \times a \times \Omega \times 
  \eta  \times f \times \mathrm{QE} \times \mathrm{CE}} \ .
\label{yield}
\end{equation}
In the following analysis we use a value of $f$ appropriate for the
main line in each filter band pass. The effect of another line in one
filter will be discussed later.  $I$ is about $(4\sim5)\times10^8$ from about
80 hours in each run.
Finally, $\epsilon$ in vacuum run is 
subtracted from each  $\epsilon$ determined above and the
corrected $\epsilon$  is determined. 
An example of typical data sample is shown in Table \ref{sample}.  

\begin{table}
\caption{An example of typical data sample. }
\begin{center}
\begin{tabular}{|l|r|r|}\hline
item  &  Nitrogen & Air \\ \hline
pressure (hPa) & 1000 & 1000 \\
filter (nm) & 337 & 337 \\

observation time (sec) & 284778 & 268919   \\
total number of electrons & 4.317$\times10^8$ & 4.106$\times10^8$  \\
coincident counts  & 34688 & 8393 \\
counts in signal portion  & 28004 & 3779 \\
signal counts & 26920 & 3148 \\
photon yield (/m$\cdot$electron) & 8.38$\pm$0.15 & 1.094$\pm$0.025  \\
photon yield in vacuum  (/m$\cdot$electron) & 0.03$\pm$0.00 & 0.028$\pm$0.001 \\
corrected photon yield  (/m$\cdot$electron) & 8.35$\pm$0.15 &
 1.066$\pm$0.025 \\ \hline
\end{tabular}
\label{sample}
\end{center}
\end{table}

\subsection{Systematic errors}
\label{subsec:systematics}

The systematic errors of the present experiment are summarized in
Table \ref{error}.  
The main uncertainty is due to the PMT
calibrations (QE and CE),
which were provided by the manufacturer (Hamamatsu Photonics). 
The CE is defined by Hamamatsu as the counts at the anode divided by the
number of photo-electrons emitted from the photocathode.
In this measurement the factory 
used a 25mm diameter area centered on the photocathode,  at a 
PMT gain setting of 5$\times10^6$ and with a discriminator
setting at $\frac{1}{3}$ of the single photon peak.
These conditions are also used in the present experiment. 
The number of photo-electrons 
was estimated in DC mode in a separate measurement. 

 The filter transmission coefficients are provided by the manufacturer
(Hi-Technology Inc.) and the error is mainly due to the various
incident angles of the photons at the filter. 
We have estimated this value from the results of M1 and M2
described in the Section 2.2 (1). 
 The error due to a
possible contamination of some lines which fall in
the tail of a filter's transmission can not be derived
accurately without absolute photon yields of those
lines.  Referring the table summarized by Bunner
we assumed these values as square root sum of six
filter bands, a few \% each ($\sqrt{6}\times 2$\%), and
is listed in the fourth line. 
Other systematic uncertainties appeared in Eq.(\ref{yield})
are as follows: I is negligible; a, 2\%; 
$\Omega$, 3\%; and $\eta$ , 2\%.
These are added in quadrature and the result is listed in
the fifth line.

It should be noted that the temperature has been maintained
within $\pm0.5^{\circ}$C, the pressure within $\pm$0.5hPa
and the high voltage within $\pm$0.05\% in each run.
The uncertainty due to the variation of these factors is
negligible in estimating $\epsilon$, but about a few\% 
in evaluating p' of 391nm line.
Therefore total systematic error in $\epsilon$
is estimated to be 14 \%, 
assuming the errors add in quadrature. 
The statistical error in each run is less than 3\%.

\begin{table}[thb]
\caption{Systematic errors of the present experiment}
\begin{center}
\begin{tabular}{|r|l|r|} \hline
No. &  item  &  errors  \\ \hline
1. & Quantum efficiency of PMT &  5 \% \\ 
2. & Collection efficiency of PMT &  10 \% \\ 
3. & Transmission coefficient of filter &    5 \%  \\ 
4. & Contamination from lines at the tail of filter transmission &  5 \%  \\ 
5. & Other parameters ($I, a, \Omega, \eta$ of Eq.(1)) &  4 \%  \\ 
 \hline
   & Total &  14 \%  \\  \hline
\end{tabular}
\end{center}
\label{error}
\end{table}

\section{Results of photon yields}
\label{sec:results}
\subsection{Nitrogen}
\label{subsec:Nitrogen}
\begin{figure} [thb]
\centerline{\includegraphics[height=15.0cm]{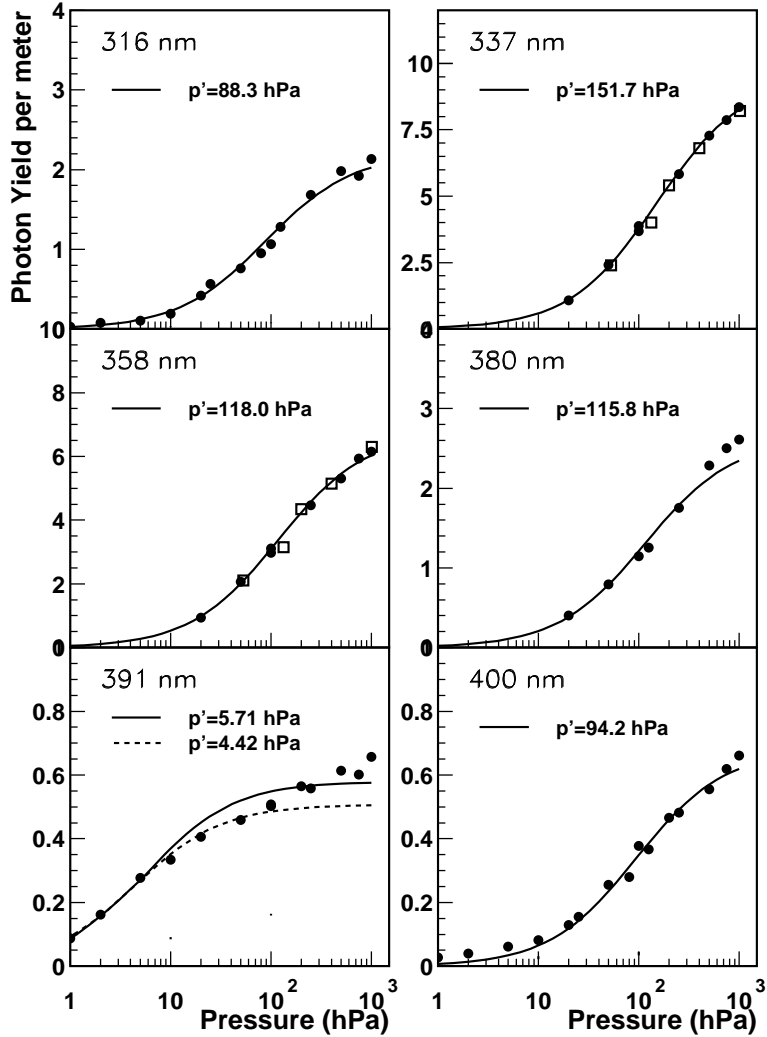}} 
\caption{The pressure dependence of $\epsilon$ in N$_2$ at
17 $^{\circ}$C excited by electrons with an average energy of 0.85 MeV
in six wave bands. In each figure, the main contributing line in each
wave band is listed.  Solid circles are the present results.  Open
squares are from Kakimoto et al. cited from Ueno \cite{ueno96}.  Solid curves
are the best fits of Eq.(\ref{eq-p}) with $p'$ as shown, discussed below in
Section \ref{subsec:decay_n2}.
The dotted line for the 391 nm band shows the best fit
when the five highest points are excluded.}
\label{nitrogen}
\end{figure}

As a first step we have measured $\epsilon$ in nitrogen, so that we
may compare results with previous experiments.  The pressure
dependencies of $\epsilon$ in six wave bands are indicated by solid
circles in Fig. \ref{nitrogen}.  In each figure, the main line in each
filter band is indicated.  The results of Kakimoto et al. (cited from
Ueno \cite{ueno96}) are also plotted with open squares in the 337 nm
and 358 nm bands.  The present results at 337 nm and 358 nm are in
good agreement with those by Kakimoto et al., even though the PMTs,
filters and gas chambers are different in the two experiments.

The solid lines in the figure are from Eq.(\ref{eq-p}), which will be
discussed in Section \ref{subsec:decay_n2}.
A dotted line in the plot for the 391 nm
band is the best fit to the equation with the five highest pressure
points excluded.  The experimental points in some filters cannot be
fitted with a constant $p'$ throughout the pressure range, especially
in the 391 nm band.  This may be due to the superposition of two lines
in one filter band, as will be discussed in Section \ref{subsec:2component}.

\subsection{Air}
\label{subsec:Air}
We used
a mixture of 78.8 \% nitrogen gas and 21.2 \% oxygen gas,
without Argon and the other small contributions to air, such as CO$_2$,
Ne, CO, water vapour etc.  This air-like mixture is expected to give
similar photon yields to dry air, since the change of photon yields may be
negligibly small even if Argon of 1 \% is included \cite{bun64}.  In
Kakimoto et al, a mixture of nitrogen and oxygen gas was also used as
dry air.

\begin{figure} [t] 
\centerline{\includegraphics[height=15.0cm]{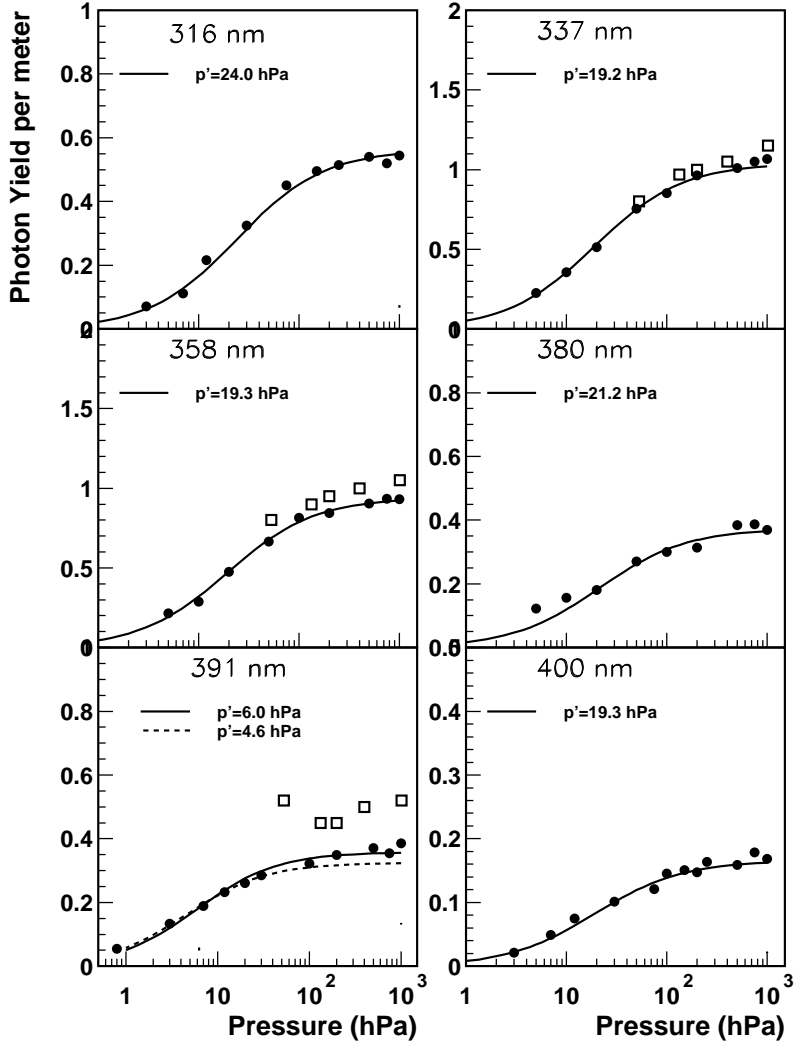}} 
\caption{The pressure dependence of $\epsilon$ in air at
20 $^{\circ}$C.  The data of Kakimoto et al. in dry air at
15 $^{\circ}$C with 1.4 MeV electrons is plotted by open squares.  Solid
lines show the best fit of Eq.(\ref{eq-p}) with the value of $p'$ as shown,
as discussed in Section \ref{subsec:decay_n2}. The dotted line in the plot of
the 391 nm dependence is the best fit excluding the four highest
points.}
\label{air}
\end{figure}

The pressure dependence of $\epsilon$ in dry air is shown in
Fig. \ref{air} by filled circles in six wave bands.  The results of
Kakimoto et al. \cite{kak96} are also shown by open squares in the 337
nm, 358 nm and 391 nm bands.  Their results are larger than the
present ones by about 0.1 photons/m irrespective of the pressure, and
in all three bands.  This is understood as follows: there were
backgrounds of photons from Bremsstrahlung at the foil of the source
window in Kakimoto et al., while those photons were reduced by a lead
brick in this experiment.  The effects of this background were
negligibly small in the nitrogen measurements as shown in
Fig. \ref{nitrogen}.  The solid lines come from Eq.(\ref{eq-p}), which will be
discussed in Section \ref{subsec:decay_air}.

\section{Analysis}
\label{sec:analysis}
\subsection{Fluorescence decay time and efficiency in N$_2$ gas} 
\label{subsec:decay_n2}
The observed decay time $\tau$ of the fluorescence is related to the
following three terms: the mean lifetime of the excited state for
decay to any lower state $\tau_r$, the lifetime of collisional
de-excitation $\tau_c$ and that of internal quenching $\tau_i$.  This can
be expressed as follows\cite{bun64},
\begin{equation}
 \frac{1}{\tau} = 
  \frac{1}{\tau_r} + \frac{1}{\tau_c} + \frac{1}{\tau_i} 
  = \frac{1}{\tau_o} + \frac{1}{\tau_c} \ ,
\label{eq-tau}
\end{equation}
and
\begin{equation}
   \tau_c =   \frac{\sqrt{\pi MkT}}{4\sigma_{nn}} \frac{1}{p} \ ,
\label{eq-tau-c}
\end{equation}
where $M$ is the N$_2$ molecular mass, $k$ is Boltzmann's constant, $T$
is the gas temperature, $\sigma_{nn}$ is the cross-section for
nitrogen-nitrogen collisional de-excitation and $p$ is the gas
pressure.  We define the reference pressure $p'$ \cite{bun64} , which
corresponds to the pressure where $\tau_o$ is equal to $\tau_c$, as
\begin{equation}
   p' = \frac{\sqrt{\pi MkT}}{4\sigma_{nn}} \frac{1}{\tau_o} \ .
\label{eq-p-dash}
\end{equation}
Then the reciprocal of the lifetime $\frac{1}{\tau}$ is related to $p$ by

\begin{equation}
  \frac{1}{\tau} = \left(\frac{1}{\tau_o p'}\right) p + \frac{1}{\tau_o}
  =\frac{1}{\tau_0}(ap+1) \ ,
\label{eq-inv-tau}
\end{equation}
where $a=\frac{1}{p'}$.

The fluorescence efficiency of the $i$-th band at pressure $p$,
$\Phi_i(p)$, is defined as the power radiated by the gas in the $i$-th
band per unit power deposited in the gas by an electron.  This
$\Phi_i(p)$ can be written as the ratio of de-excitation by radiation
to the total, and is expressed as a function of $p$ as follows
\cite{bun64}:
\begin{equation}
 \Phi_i(p) =  \frac{\frac{1}{\tau_r}}{\frac{1}{\tau_o}+\frac{1}{\tau_c}}
	=  \frac{\frac{\tau_o}{\tau_r}}{1 + \frac{\tau_o}{\tau_c}}
	=  \frac{\frac{\tau_o}{\tau_r}}{1 + \frac{p}{p'_i}} \ .
\label{eq-eff}
\end{equation}
Since the energy available to produce photons in the $i$-th band is
the electron energy loss per unit length times $\Phi_i(p)$, the photon yield
per unit length per electron, $\epsilon_i$, for $i$-th band under 
atmospheric pressure $p$ is written as follows \cite{elb93}:
\begin{equation}
\epsilon_i = \rho \left(\frac{\d E}{\d x}\right) 
\left(\frac{\Phi_i(p)}{h \nu_i}\right) \ , \\
\label{epsilon-i-engy}
\end{equation}
and 
\begin{equation}
\Phi_i(p) = \frac{\Phi_i^{\circ}}{1+\frac{p}{p'_i}} \ ,
\label{phi-i}
\end{equation}
where $\rho$ is the air density (kg m$^{-3}$), $\displaystyle
\frac{\d E}{\d x}$ is the energy loss in (eV kg$^{-1}$ m$^{2}$), and $h
\nu_i$ is the photon energy of the $i$-th band (eV).  $\Phi_i^{\circ}$
(=$\frac{\tau_o}{\tau_r}$) corresponds to the fluorescence efficiency
for the $i$-th band in the absence of collisional quenching.

By using the equation of state of a gas, $p=\rho R_{\mathrm{N}_2} T$ where
$R_{\mathrm{N}_2}$ is the specific gas constant in N$_2$ (296.9
m$^2$s$^{-2}$K$^{-1}$), Eq.(\ref{epsilon-i-engy}) is rewritten as a
function of $p$ at a constant temperature $T$ in Kelvin:
\begin{eqnarray}
\epsilon_i &=& \frac{p}{R_{\mathrm{N}_2}T h\nu_i}\left(\frac{\d E}{\d x}\right) 
\label{eq-pres}
\left(\frac{\Phi_i^{\circ}}{1 + \frac{p}{p'_i}} \right) \\
           &=& \frac{C_i p}{1 + \frac{p}{p'_i}} 
            = \frac{C_i}{x + a_i} \ ,
\label{eq-p}
\end{eqnarray}
where $x=\frac{1}{p}$ and $a_i=\frac{1}{p'_i}$.  This relation can be written in a more general form 
as a function of the gas density and the temperature as \cite{kak96}
\begin{equation}
\epsilon_i = \frac{A_i \rho}{1 + \rho B_i \sqrt{T}} \ , 
\label{ep-temp}
\end{equation}
where 
\begin{equation}
 A_i  = \frac{\left(\frac{\d E}{\d x}\right) \Phi_i^{\circ}}{h \nu_i} 
 \ \ \ \ \mathrm{and}
 \ \ \ \  B_i = \frac{R_{\mathrm{N}_2}\sqrt{T}}{p'_i}=
	\frac{4\sigma_{nn}\tau_o R_{\mathrm{N}_2}}{\sqrt{\pi k M}} \ .
 \nonumber
\end{equation}

\begin{figure}[thb]
\centerline{\includegraphics[height=8cm]{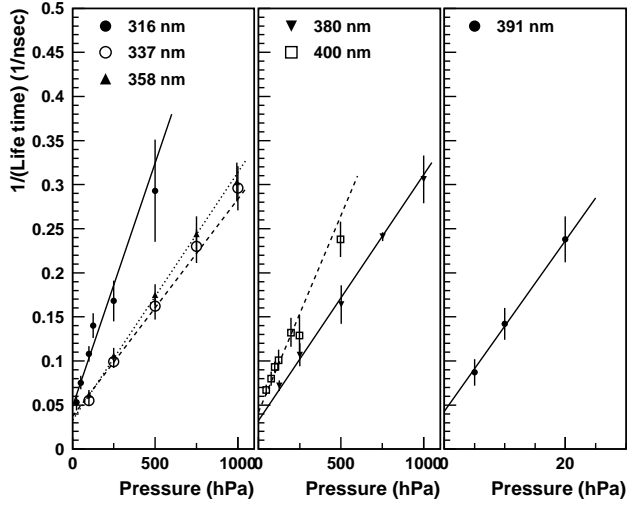}} 
\caption{The pressure dependencies of $\frac{1}{\tau}$
 for the excited states of N$_2$ for 6 wave bands.   
}
\label{inv_tau}
\end{figure}

The pressure dependencies of $\frac{1}{\tau}$ for various wave bands 
in N$_2$  are shown in Fig. \ref{inv_tau}. 
 Each $\tau$ and its standard deviation in the figure was
calculated with a weighted least square by
fitting the counts($J$) in the decay portion of the
TDC distribution (see Fig. \ref{distribution}) to the
equation $\ln J=\ln J_0 - \frac{t}{\tau}$. Due to the
time resolution of the present experiment (1.26nsec) and
the limited statistics, the error in $\frac{1}{\tau}$ of each point 
is quite large compared with that of $\epsilon$.

In order to fit the experimental points of Fig. \ref{nitrogen} and
\ref{inv_tau} to Eq.(\ref{eq-p}) and (\ref{eq-inv-tau}) respectively,
 the least square (LS) method is utilized.  Let us consider first
the fit to the $p-\epsilon$ diagram (Fig. \ref{nitrogen}),
 since the data points are
more numerous than those in $p-\frac{1}{\tau}$ diagram
(Fig. \ref{inv_tau}) and the large difference of errors
between both diagrams as mentioned above. 
For each $i$-th band
the first $\chi^2_1$ is defined by 
\begin{equation}
\chi^2_1=\sum_j \omega_j(\epsilon_j-\frac{C}{x_j+a})^2 \ ,
\end{equation}
where the weights $\omega_j$ come from the errors in $\epsilon_j$ at
$p_j(=\frac{1}{x_j})$.  By minimizing $\chi^2_1$, we find the first
values of $a^{(1)}$ and $C^{(1)}$.  Then after linearizing
Eq.(\ref{eq-p}) using $a^{(1)}$ and $C^{(1)}$, Newton's method
\cite{fro79} is applied to estimate the final $a^{(\nu)}$ and
$C^{(\nu)}$ after the $\nu$-th iteration.  The statistical 
errors of $a^{(\nu)}$
and $C^{(\nu)}$ are evaluated using the propagation law of errors
using those of $\epsilon_j$.  For a given $a^{(\nu)}$, it is easy to
find $\tau_0$ and its error by the LS method because
Eq.(\ref{eq-inv-tau}) is linear with respect to $(ap+1)$.  The
obtained results are shown by solid curves in Fig. \ref{nitrogen} and
by lines in Fig. \ref{inv_tau}.  The parameters are listed in Table
\ref{tau_value}.  The values of $\sigma_{nn}$ derived from $\tau_0 p'$
and $\Phi^{\circ}$ from $C$ are also listed in the table.  In the case
of the 391 nm band, we find that the experimental points do not fit
well to one value of $p'$ throughout the pressure range but a dashed
line in Fig. \ref{nitrogen} shows Eq.(\ref{eq-p}) for a limited range
of pressures.

\begin{table}[thb]
\caption{$\tau_o$, $p'$, $\sigma_{nn}$, $C$ and $\Phi^{\circ}$ 
 in  N$_2$.}
\begin{center}
\catcode`?=\active \def?{\phantom{0}}
\begin{tabular}{|c|c|c|c|c|c|} \hline
wavelength &  $\tau_o$  & $p'$  & $\sigma_{nn}$ &  $C$ & $\Phi^{\circ}$ \\ \hline
 nm & \multicolumn{1}{|c|}{ns} & hPa & 
$\times10^{-20} $m$^2$ &  $\times10^{-2}$ /(hPa$\cdot$m) & $\times10^{-3}$ 
\\ \hline
316 & 20.5$\pm$?2.2 & ?88.3?$\pm$?7.5? & ?3.36$\pm$?0.46 & 
?2.50$\pm$0.14 & 0.509$\pm$0.028 \\ 
337 & 26.8$\pm$?1.4 & 151.8?$\pm$?4.8? & ?1.50$\pm$?0.09 &
?6.27$\pm$0.13 & 1.20?$\pm$0.03? \\ 
358 & 30.2$\pm$?0.6 & 118.0?$\pm$?4.0? & ?1.71$\pm$?0.07 &  
?5.71$\pm$0.13 & 1.03?$\pm$0.02? \\ 
380 & 30.9$\pm$?0.6 & 115.9?$\pm$11.1? & ?1.70$\pm$?0.17 & 
?2.26$\pm$0.10 & 0.382$\pm$0.017 \\ 
391 & 23.4$\pm$?0.6 & ??4.42$\pm$?0.28 & 59.0?$\pm$?4.1? &
11.5?$\pm$0.5? & 1.90?$\pm$0.09? \\ 
400 & 23.8$\pm$?1.8 & ?94.2?$\pm$?7.9? & ?2.72$\pm$?0.31 & 
?0.72$\pm$0.04 & 0.116$\pm$0.007 \\ \hline
\end{tabular}
\end{center}
\label{tau_value}
\end{table}

\begin{table}[thb]
\caption{Comparison of $\Phi_i(p)$ in  N$_2$ at
$p$=800 hPa (600 mmHg). D. and O. refers to Davidson and O'Neil \cite{dav64}.}
\begin{center}
\catcode`?=\active \def?{\phantom{0}}
\begin{tabular}{|c|c|c|c|} \hline
    & \multicolumn{1}{|c|}{This experiment} & Ueno
& D. \& O.  \\ \hline
beam &  \multicolumn{1}{|c|}{electron} & electron & electron \\ \hline
energy & \multicolumn{1}{|c|}{0.85 MeV}& 1.4 MeV & 50 keV    \\ \hline
wavelength & \multicolumn{3}{|c|}{$\times10^{-4}$} \\ \hline
316 nm &    0.505$\pm$0.051  &               &       \\ 
337 nm &    1.90?$\pm$0.07?  & 1.87$\pm$0.01 & 5.20?   \\ 
358 nm &    1.32?$\pm$0.05?  & 1.35$\pm$0.01 & 3.70?   \\ 
380 nm &    0.483$\pm$0.051  &               & 1.40?   \\ 
391 nm &    0.104$\pm$0.007  &               & 0.102   \\ 
400 nm &    0.122$\pm$0.013  &               & 0.200   \\ \hline 
\end{tabular}
\end{center}
\label{comp_n2}
\end{table}

Our values of $\Phi_i(p)$ in N$_2$ at 800 hPa are listed in Table
\ref{comp_n2}, together with those of other experiments.  The electron
energy loss $\frac{\d E}{\d x}$ is calculated using the equation in
Sternheimer et al. \cite{ster82} with a density correction, and is
0.1687 MeV/kg$\cdot$m$^{-2}$ and 0.1677 MeV/kg$\cdot$m$^{-2}$ at 0.85
MeV, for N$_2$ and air respectively.  The values from the experiments
by Kakimoto et al. are cited from Ueno \cite{ueno96} and agree well
with the present ones for the 337 nm and 358 nm bands.  However, the
values by Davidson \& O'Neil \cite{dav64} are about a factor of 2.8
larger than the present results except at 391 nm and 400 nm.

The present result for $\Phi^{\circ}_i$ at 391 nm is
(1.90$\pm$0.09)$\times10^{-3}$ which does not agree with the
result (6.0$\pm 1.9)\times 10^{-3}$ by M.N.Hirsh et
al. \cite{hir70}, measured using electrons of 1.46 MeV under a
gas pressure of less than 10 hPa.  This will be discussed in
Section \ref{subsec:2component} in connection with the two lines 
contained within this filter band.

\subsection{Fluorescence decay time and efficiency in dry air}
\label{subsec:decay_air}

In the case of air, the reference pressure $p'$ 
of Eq.(\ref{eq-p-dash}) may be written as \cite{bun64}:
\begin{equation}
\frac{1}{p'} = \frac{4 \tau_{\circ}}{\sqrt{\pi M_n k T}} 
\left(f_n \sigma_{nn} + f_o \sigma_{no}\sqrt{\frac{M_n + M_o}{2M_o}}
\right) = \frac{D}{\sqrt{T}} \ ,
\label{ref-pres}
\end{equation}
where  $M_n$ and $M_o$ are the 
masses of nitrogen and oxygen molecules, respectively.
$\sigma_{nn}$ and $\sigma_{no}$ 
are the cross-sections for collisional loss by excited nitrogen 
molecules without radiation, through collisions with other nitrogen 
or oxygen molecules, respectively. 
The fraction of nitrogen $f_n$ is 0.79 and 
that of oxygen $f_o$ is 0.21.

\begin{figure}
\centerline{\includegraphics[height=8cm]{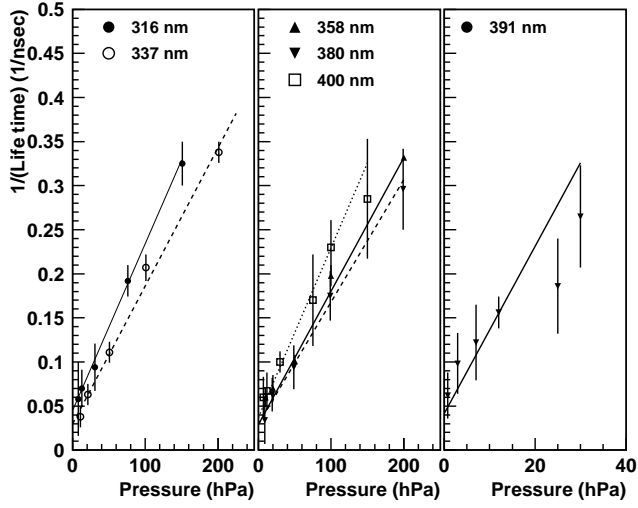}} 
\caption{The pressure dependence of the  reciprocal
 of the fluorescence decay time ($\frac{1}{\tau}$) in air.}
\label{inv_tau_air}  
\end{figure}

\begin{table}[thb]
\caption{$\tau_o$, $p'$, $\sigma_{no}$, $C$ and $\Phi^{\circ}$ 
 in  air}
\begin{center}
\catcode`?=\active \def?{\phantom{0}}
\begin{tabular}{|c|c|c|c|c|c|} \hline
wavelength &  $\tau_o$  & $p'$  & $\sigma_{no}$ &  $C$ & $\Phi^{\circ}$ \\ \hline
 nm & \multicolumn{1}{|c|}{ns} & hPa & 
$\times10^{-19} $m$^2$ &  $\times10^{-2} $/(hPa$\cdot$m) & $\times10^{-3}$ 
\\ \hline
316 & 22.0$\pm$0.5 & 24.0?$\pm$2.0? & 4.30$\pm$0.70 & 
 2.34$\pm$0.14 & 0.463$\pm$0.029 \\ 
337 & 32.4$\pm$1.8 & 19.2?$\pm$0.7? & 4.06$\pm$0.29 &
 5.43$\pm$0.15 & 1.01?$\pm$0.03? \\ 
358 & 34.2$\pm$0.4 & 19.3?$\pm$1.4? & 3.83$\pm$0.31 &  
 4.86$\pm$0.27 & 0.850$\pm$0.047 \\ 
380 & 33.9$\pm$1.6 & 21.2?$\pm$2.6? & 3.45$\pm$0.46 & 
 1.76$\pm$0.18 & 0.289$\pm$0.029 \\ 
391 & 24.2$\pm$3.2 & ?4.61$\pm$0.34 & 3.70$\pm$0.62 &
 7.04$\pm$0.39 & 1.13?$\pm$0.06? \\ 
400 & 26.9$\pm$1.9 & 19.3?$\pm$1.8? & 4.65$\pm$0.76 & 
 0.86$\pm$0.06 & 0.134$\pm$0.009 \\ \hline
\end{tabular}
\end{center}
\label{tau_air}
\end{table}

The pressure dependence of $\frac{1}{\tau}$ in dry air is shown
in Fig. \ref{inv_tau_air} for six wave bands.  The fitting of
parameters from Eq.(\ref{eq-p}) and (\ref{eq-inv-tau}) to the
data points in Fig. \ref{air} and \ref{inv_tau_air} is done using
the same method as described for the N$_2$ case.  Three
parameters, $a$ (or $p'$), $C$ and $\tau_o$ and their errors are
thus evaluated.  The results are shown by solid curves in
Fig. \ref{air} and by lines in Fig. \ref{inv_tau_air} 
and the parameters are
listed in Table \ref{tau_air}.  
 Since $p'$ is determined from the pressure dependence
of $\epsilon$, the line of 391nm in Fig. \ref{inv_tau_air}
seems not to be best fitted to the experimental points.  
The cross-section $\sigma_{no}$
is derived from $\tau_0 p'$, where $p'$ is given by
Eq.(\ref{ref-pres}) with $\sigma_{nn}$ determined previously.
$\Phi^{\circ}$ and $C$ are also included in the table.  In the
case of 391 nm, the dashed curve in Fig. \ref{air}
represents parameters in Table \ref{tau_air}.

\begin{table}[thb]
\caption{Comparison of $\Phi_i(p)$ in air at $p$=800 hPa}
\begin{center}
\catcode`?=\active \def?{\phantom{0}}
\begin{tabular}{|c|c|c|c|c|c|} \hline
    & \multicolumn{1}{|c|}{This experiment} & Kakimoto
& D. \& O. & Hartman & Bunner \\ \hline
beam &  \multicolumn{1}{|c|}{electron} & electron & electron &
electron & $\alpha$ \\ \hline
energy & 0.85 MeV  & 1.4 MeV & 50 keV &  &  4 MeV \\ \hline
nm  & \multicolumn{5}{|c|}{$\times10^{-6}$} \\ \hline
316 & 13.1?$\pm$1.4? &      &              & ?3.0  & 5.3 \\ 
337 & 23.7?$\pm$1.1? & 21?? & 21.0$\pm$3.2 & 20.0  & 8.5 \\ 
358 & 20.2?$\pm$1.8? & 22?? & 15.0$\pm$2.3 & 16.0  & 5.8 \\ 
380 & ?7.5?$\pm$1.2? &      & ?5.2$\pm$0.8 &       & 4.1 \\ 
391 & ?6.48$\pm$0.60 & ?8.4 & ?7.0$\pm$1.1 & ?5.9  & 4.8 \\ 
400 & ?3.18$\pm$0.36 &      & ?1.8$\pm$0.3 & ?1.7  & 1.5 \\ \hline 
\end{tabular}
\end{center}
\label{comp_air}
\end{table}

The yield in air $\epsilon_i$ is expressed by Eq.(\ref{ep-temp}),
after replacing $R_{\mathrm{N}_2}$ by $R_{\mathrm{air}}$=287.1
m$^2$s$^{-2}$K$^{-1}$ and by using $p_i'$ given by
Eq.(\ref{ref-pres}).  The values of $\Phi_i(p)$ in air at 800 hPa
are listed in Table \ref{comp_air} together with results of other
experiments.  Those of Hartman and Bunner are calculated by using
$\Phi_i^{\circ}$ and $p'$ listed in Table 2.6 of Bunner
\cite{bun64}.  These values of $\Phi_i(p)$ at $p$=800 hPa are in
good agreement, except those of Bunner which were measured using
alpha beams.

\begin{figure}[thb]
\centerline{\includegraphics[height=8cm]{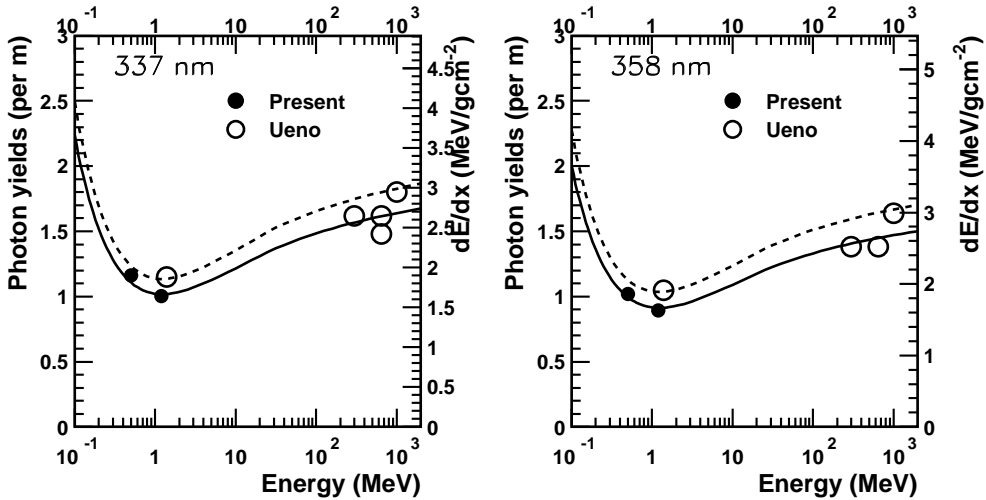}}
\caption{Energy dependence of $\epsilon$ in air at 1000 hPa. 
The $\d E/\d x$ curve is shown by a solid line, and is normalized to the yield
at 1.2 MeV (present measurement).}
\label{de-dx}
\end{figure}

In order to see the energy dependence of $\epsilon_i$ in air,
values of $\epsilon_i$ are derived in two energy regions below
and above the average energy (0.85 MeV).  In Fig. \ref{de-dx},
the energy dependence of $\epsilon_i$ in air at 1000 hPa is shown
for two wave bands together with measurements by Kakimoto et
al. cited from the thesis of Ueno \cite{ueno96}.  The
$\frac{\d E}{\d x}$ of an electron in air is calculated from a
formula given in Sternheimer et al.\cite{ster82} and the solid
line is normalized to the 1.2 MeV yield point of the present
measurements.  A dashed line is normalized to the 1.4 MeV point
of Ueno in each figure.  The solid lines agree with the higher
energy measurements (between 300 MeV and 1000 MeV) better than
the dashed lines.  This may be because the photon backgrounds
were not excluded from the 1.4 MeV points of Ueno, as described
in Section \ref{subsec:Air}.  Our measurements are consistent
with $\epsilon$ being
proportional to $\frac{\d E}{\d x}$.

\subsection{Two components analysis in one filter band}
\label{subsec:2component}

There are some discrepancies between the experimental points and the
model pressure dependencies of $\frac{1}{\tau_i}$ and $\epsilon_i$ as
shown in Eqs.(\ref{eq-inv-tau}) and (\ref{eq-p}), especially in the 391 nm
filter band.  In the following, we try to fit the observed
pressure dependencies of $\frac{1}{\tau_i}$ and $\epsilon_i$ with
a superposition of two lines in one filter band.  In this case
the observed photon yield $\epsilon_{\mathrm{obs}}(p)$ is the sum of the
photon yields of the main line $\epsilon_{1}(p)$ and the sub-line
$\epsilon_{2}(p)$, and is written as follows by extending Eq.(\ref{eq-p}):

\begin{eqnarray}
\epsilon_{\mathrm{obs}}(p) = \epsilon_1(p) + \epsilon_2(p) 
             &=& \frac{C_1}{x+a_1} 
           + \frac{C_2}{x+a_2} \ \ , 
\label{epsilon-two}
\end{eqnarray}
where $x=\frac{1}{p}$, $a_1=\frac{1}{p'_1}$ and
$a_2=\frac{1}{p'_2}$.  $C_1$ and $p'_1$ are the parameters of the
main line in the filter band, and $C_2$ and $p'_2$ are the parameters
of another line.  It should be noted that $\epsilon_{\mathrm{obs}}$ is
determined assuming the filter transmission of the main line in
the filter, and hence $\epsilon_2$ must be corrected with using
the filter transmission of that line.

The reciprocal of the observed life time
$\frac{1}{\tau_{\mathrm{obs}}(p)}$ is approximately expressed by the
weighted mean of $\frac{1}{\tau_{1}(p)}$ and
$\frac{1}{\tau_{2}(p)}$.  The weights are expressed in terms of the
relative photon intensities of two lines.  Therefore,
\begin{eqnarray}
 \frac{1}{\tau_{\mathrm{obs}}(p)} &=&
  \left(\frac{\epsilon_1(p)}{\epsilon_1(p) + \epsilon_2(p)}\right)
   \frac{1}{\tau_1(p)}
 + \left(\frac{\epsilon_2(p)}{\epsilon_1(p) + \epsilon_2(p)}\right)
   \frac{1}{\tau_2(p)}  \  , 
\label{two-life}
\end{eqnarray}
where
\begin{eqnarray}
 \frac{1}{\tau_1(p)}&=&\frac{1}{\tau_{o1}}(a_1 p + 1)  \ ,   \nonumber \\
 \frac{1}{\tau_2(p)}&=&\frac{1}{\tau_{o2}}(a_2 p + 1)  \ . 
\label{two-tau}
\end{eqnarray}

In this case, we have to determine a set of four parameters
$a_1$, $C_1$, $a_2$ and $C_2$ in Eq.(\ref{epsilon-two}).  For
many such sets, the value of $\chi^2$ is computed and the set
having the minimum $\chi^2$ is taken as the initial set.
Newton's method is also applied for this set, which leads to the
final set after the $\nu$-th iteration.  Values of
$a_1(=\frac{1}{p'_1})$, $C_1$, $a_2(=\frac{1}{p'_2})$ and $C_2$
determined in this way are listed in the upper half of Tables
\ref{n2_two} and \ref{air_two}.
 The values of minimum $\chi^2$ are also depicted as 
$\chi^2$ in the tables.

The errors of these four parameters (as listed in the tables) are
estimated in the following way.  Since the probability density
function (pdf) of $\Delta\chi^2$, i.e. the deviation from
$\chi^2_{\mathrm{min}}$, follows the pdf of $\chi^2$ with four degrees of
freedom, we take $\Delta\chi^2_{\mathrm{critical}}$=4.72 where
the coverage probability is 0.683 corresponding to the
1-sigma \cite{par02}.  Many sets of ($p''_1$, $C_1'$,
$p''_2$, $C_2'$) near to ($p'_1$, $C_1$, $p'_2$, $C_2$) are
selected such that the value of $\chi^2$ almost equals
$\chi^2_{\mathrm{min}}+\Delta\chi^2_{\mathrm{critical}}$.  
The maximum and minimum
values of $(p''_1 - p'_1)$ lead to the error bounds of $p'_1$.
Errors of other three parameters are estimated in the same way.

From Eq.(\ref{epsilon-two}), (\ref{two-life}) and (\ref{two-tau}),
$\frac{1}{\tau_{\mathrm{obs}}(p)}$ is expressed as
\begin{eqnarray}
 \frac{1}{\tau_{\mathrm{obs}}(p)}=\frac{p}{\epsilon_1(p)+\epsilon_2(p)}
   \times v \ \ , 
\label{two-tau-fit}
\end{eqnarray}
where 
\begin{eqnarray}
 v=\frac{C_1}{\tau_{o1}}+\frac{C_2}{\tau_{o2}} \ \ . 
\label{two-tau-fit-v}
\end{eqnarray}
That is, $\tau_{o1}$ and $\tau_{o2}$ can't be evaluated independently
without any assumptions and 
only the value $v$ 
is determined by the LS method.
It should be noted that $v$ is almost insensitive to $p'$ and C, since
these parameters are determined to fit $\epsilon_{obs}$.

\begin{figure}[thb]
\centerline{\includegraphics[height=6cm]{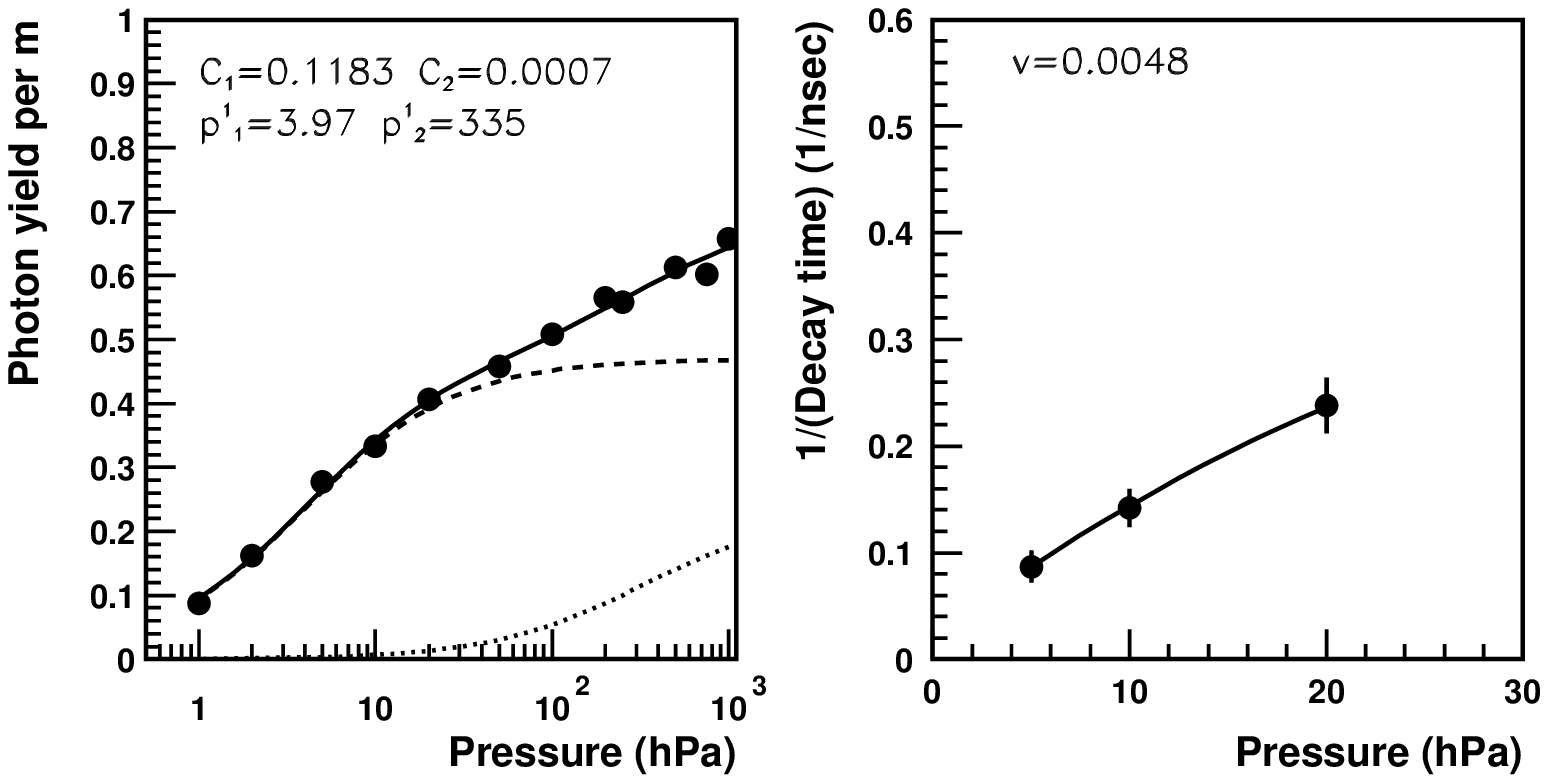}}
\centerline{\includegraphics[height=6cm]{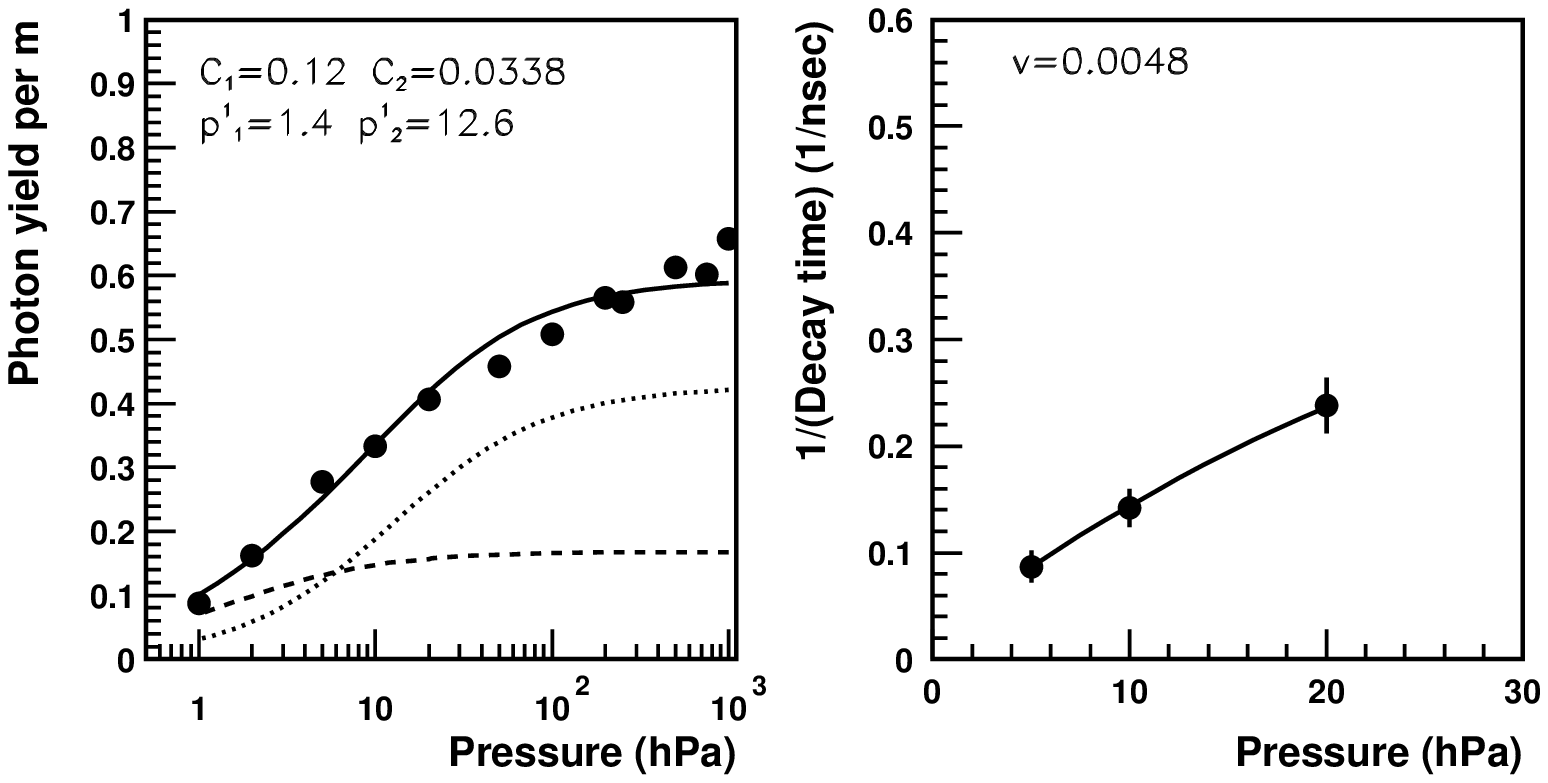}}
\caption{Two line fitting for the 391 nm bands in N$_2$.  The
upper figure is the best fit combination of two lines.  The lower
figure is the best fit for a fixed $p'_1$.  The contribution from
the main line is shown by a dashed curve, and the secondary line
by a dotted curve.  The solid curves are the sum of two lines in
the left-hand figures and the solid line is  
 Eq.(\ref{two-tau-fit}) with $v$ listed in the right-hand figures.}
\label{two_line_n2}
\end{figure}

\begin{table}[thb]
\caption{Parameters of two line fitting in the 391 nm band of N$_2$.}
\begin{center}
\catcode`?=\active \def?{\phantom{0}}
\begin{tabular}{|c|c|c|c|c|c|} \hline
Wave length &  $p'$  & $v$ &  $C$
 &$\Phi^{\circ}$ & $\chi^2$\\ \hline
nm &  hPa & $10^{-3}$/(hPa$\cdot$m$\cdot$ns) 
& $10^{-2}$/(hPa$\cdot$m) & 10$^{-3}$ & \\ \hline 
 391 &  ??3.97$\pm$0.37 & 4.79$\pm$0.36 & 11.8??$\pm$0.7?? & 1.94?$\pm$0.11? & ?32.7 \\ 
 394 &  335.??$^{+242}_{-102}$ &   
                               & ?0.071$\pm$0.028 & 0.012$\pm$0.005 & \\ \hline
 391 &  ??1.40?????     & 4.79$\pm$0.36 & 12.0??$\pm$2.0?? & 1.95?$\pm$0.33? & 158.? \\ 
 394 &  ?12.6?$^{+7.1}_{-1.0}$ &  & ?3.38?$\pm$0.26? & 0.55?$\pm$0.04? & \\ \hline
\end{tabular}
\end{center}
\label{n2_two}
\end{table}

The best fit result for a superposition of two lines in the 391 nm
band of N$_2$ is shown in the upper figure of
Fig. \ref{two_line_n2}, where the contribution from the main line
is shown as a dashed curve and that from the second line is shown
as a dotted curve.  The solid curve in the left-hand figure is the sum of
the dashed and dotted lines. In the right-hand
figure, a solid line is drawn with the value $v$.
 The parameters of each line are listed in
Table \ref{n2_two}.

It is clear that the fitting is much better than the single line
fitting in Fig. \ref{nitrogen}.  However, the value of
$\Phi^{\circ}$ at 391 nm is much smaller than the measurement of
Hirsh et al.\cite{hir70}, which was determined using 1.45 MeV
electrons in the pressure range between 0 and 10 hPa.  If we fit
their pressure dependence of photon yields with Eq.(\ref{eq-p}) and with
fixed $p_1'$=1.40 hPa, the best fit case is shown in the lower
half of Fig. \ref{two_line_n2} and the parameters are listed in
the lower half of Table \ref{n2_two}.  The difference of
$\Phi^{\circ}$ at 391 nm  is within experimental errors, 
 but the difference of $\Phi^{\circ}$ from Hirsh et al. is 
still beyond the measurement
errors of both experiments.  To resolve the issue, we need to
measure the yields at 391 nm with a narrower bandwidth filter
at low pressure.

\begin{figure}[thb]
\centerline{\includegraphics[height=6cm]{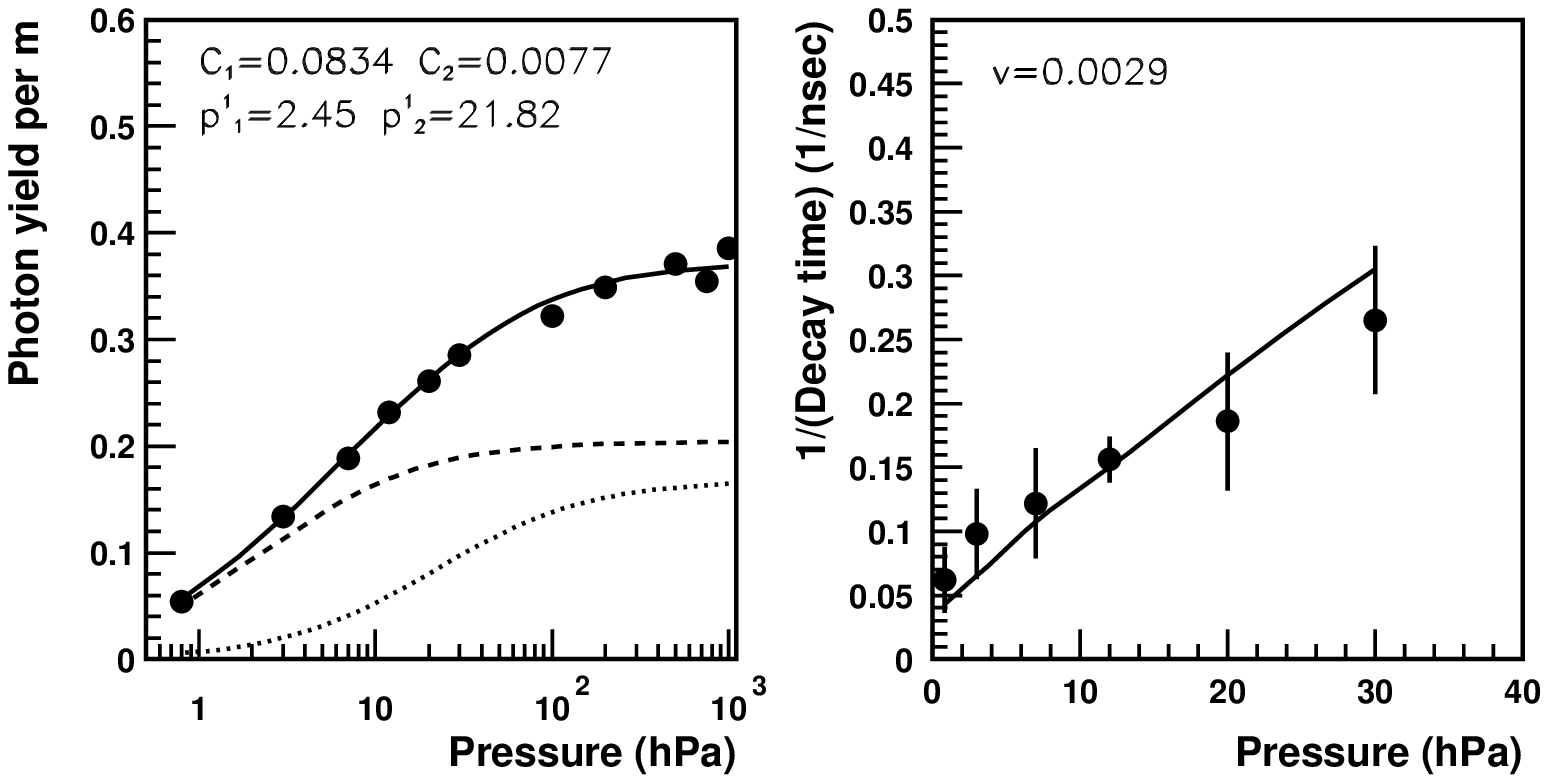}}
\centerline{\includegraphics[height=6cm]{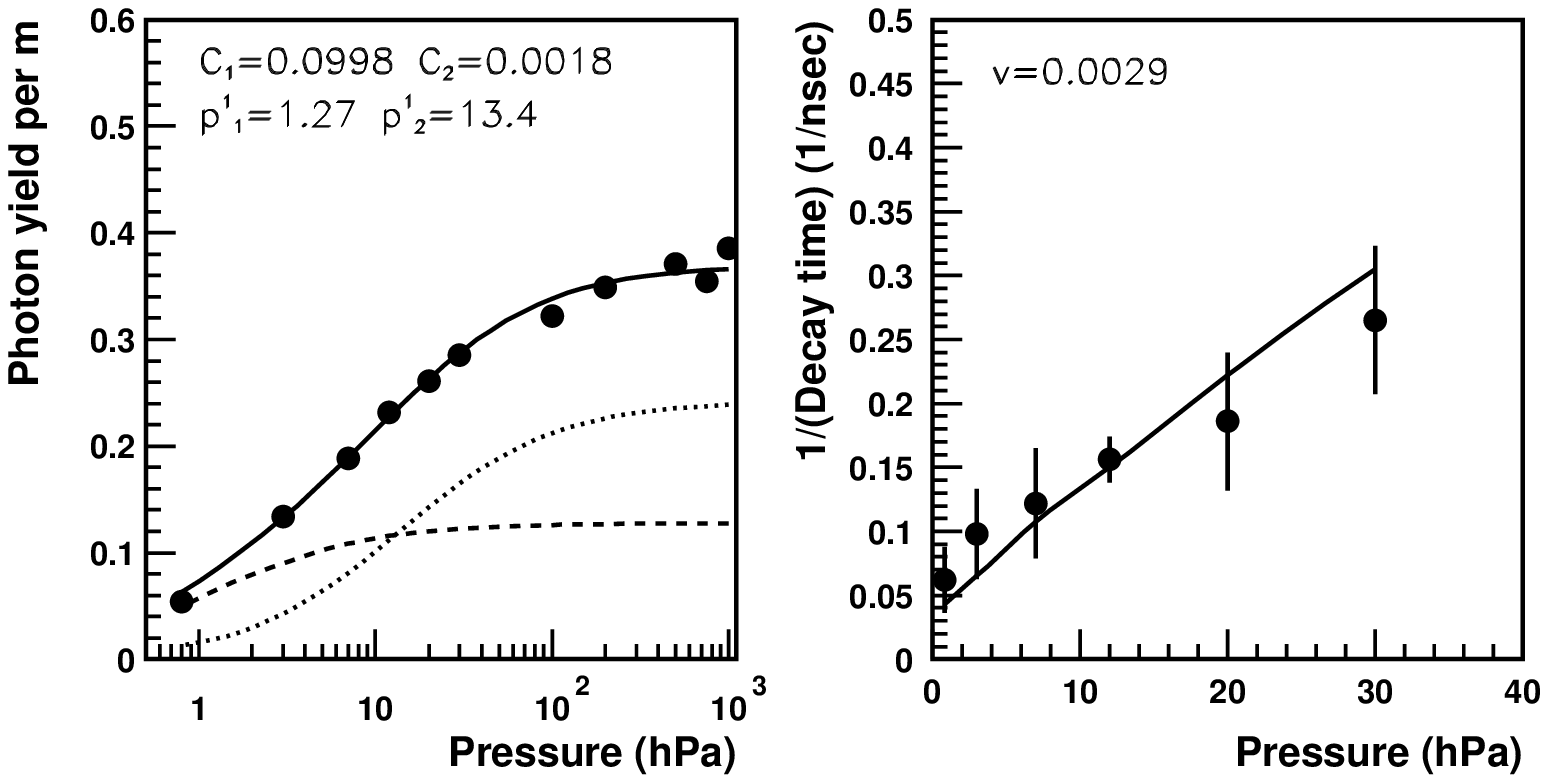}}
\caption{Two line fitting of the 391 nm band in air.  Other details are
 the same as in Fig.\ref{two_line_n2}. }
\label{two_line_air_391}
\end{figure}

\begin{table}[thb]
\caption{Parameters of two line fitting in the 391 nm band in air.}
\begin{center}
\catcode`?=\active \def?{\phantom{0}}
\begin{tabular}{|c|c|c|c|c|c|} \hline
Wave length &  $p'$  & $v$ &  $C$
 &$\Phi^{\circ}$ & $\chi^2$\\ \hline
nm  & hPa & $10^{-3}$/(hPa$\cdot$m$\cdot$ns) 
& $10^{-2}$/(hPa$\cdot$m) & 10$^{-3}$ & \\ \hline 
 391 & ?2.45$\pm$0.85 & 2.89$\pm$0.26 & ?8.34$\pm$1.44 & 1.33$\pm$0.23 & 26.8 \\ 
 394 & 21.8?$\pm$8.5? &  & ?0.77$\pm$0.33 & 0.12$\pm$0.05 &  \\  \hline
 391 & ?1.27?????     & 2.90$\pm$0.26 & ?9.98$\pm$0.14 & 1.60$\pm$0.02 & 33.0 \\ 
 394 & 13.4?$\pm$2.3? &  & ?1.80$\pm$0.35 & 0.29$\pm$0.06 & \\  \hline
\end{tabular}
\end{center}
\label{air_two}
\end{table}

In the case of the 391 nm band in air, two cases are also shown
in Fig. \ref{two_line_air_391}.  As before, the first shows the
best fit combination of two lines, allowing freedom in the
pressure dependence of each line.  The second set of plots shows
the case where we use a fixed value of $p_1'$ as determined by
Hirsh et al.\cite{hir70}.

In the other filter bands we have also determined the best fit
$p'$ and $\tau_0$ values for two lines in a filter band.
However, the best fit values of $p'$ are quite
different in some cases for the 337 nm, 356 nm, 380 nm and 400 nm bands,
where lines of the 2P(0,$v''$) and 2P(1,$v''$) transitions are
superimposed.  We have decided not to list the values in this
article and we continue the experiments with a narrower band width
and improvements of the analysis method.

\section{Discussion}
\label{sec:discussion}
\subsection{Summary of the results}
\label{subsec:summary}

\begin{table}[thb]
\caption{Summary of the present experiments. Photon yields
($\epsilon$) per meter per electron of average energy 0.85 MeV are
values in dry air at 1000 hPa and 20 $^{\circ}$C.  Values in
parentheses with an upper suffix $^a$ have not been measured in this
experiment and are estimated as described in the text.  Values in
parentheses with an upper suffix $^b$ are corrected values
taking into account the transmission coefficient (see text for
detail).}

\begin{center}
\catcode`?=\active \def?{\phantom{0}}
\begin{tabular}{|r|c|c|c|l|c|c|} \hline
i & $\lambda$ & transition & \multicolumn{1}{|c|}{$\epsilon$} 
& \multicolumn{1}{|c|}{$\Phi^{\circ}$}
 & $A$ & $B$  \\ \cline{2-2} \cline{4-7}
  & nm & state & 
 \multicolumn{1}{|c|}{m$^{-1}$} & \multicolumn{1}{|c|}{$\times10^{-3}$} 
 & m$^2$kg$^{-1}$ & m$^3$kg$^{-1}$K$^{-\frac{1}{2}}$  \\ \hline
1 & 311.7 & 2P(3,2) &        &        &       &   \\ \cline{1-3} 
2 & 313.6 & 2P(2,1) & 0.552$\pm$0.033 & 0.463$\pm$0.029 & 
19.6?$\pm$1.2? & ?2.04$\pm$0.17  \\ \cline{1-3} 
3 & 315.9 & 2P(1,0) &        &        &       &   \\ \hline
4 & 328.5 & 2P(3,3) &        &        &       &   \\ \cline{1-3}
5 & 330.9 & 2P(2,2) & (0.041$\pm$0.012)$^a$   &  & & \\ \hline
6 & 333.9 & 2P(1,1) &        &        &       &   \\ \cline{1-3}
7 & 337.1 & 2P(0,0) & 1.024$\pm$0.028 & 1.006$\pm$0.027 & 
45.4?$\pm$1.2? & ?2.56$\pm$0.09 \\ \hline
8 & 346.9 & 2P(3,4) &        &        &       &   \\ \cline{1-3}
9 & 350.0 & 2P(2,3) & (0.030$\pm$0.009)$^a$   &  & & \\ \hline
10& 353.7 & 2P(1,2) &        &        &       &   \\ \cline{1-3}
11& 357.7 & 2P(0,1) & 0.926$\pm$0.051 & 0.850$\pm$0.047 &
 40.7?$\pm$2.3? & ?2.54$\pm$0.18 \\ \hline
12& 367.2 & 2P(3,5) &        &        &       &    \\ \cline{1-3}
13& 371.1 & 2P(2,4) & (0.054$\pm$0.016)$^a$ & & &  \\ \hline
14& 375.6 & 2P(1,3) & (0.493$\pm$0.074)$^b$ &  (0.388$\pm$0.058)$^b$ 
 &  (19.8?$\pm$3.0?)$^b$  &   \\ \cline{1-3}
15& 380.5 & 2P(0,2) & 0.367$\pm$0.038 & 0.289$\pm$0.029 &
 14.7?$\pm$1.5? & ?2.31$\pm$0.25  \\ 
\hline
17& 391.4 & 1N(0,0) & 0.204$\pm$0.079 & 1.33?$\pm$0.23? & 
 69.8?$\pm$12.? & 20.1?$\pm$6.9?  \\ \hline
16& 389.4 & 2P(3,6) &        &        &       &  \\ \cline{1-3} 
18& 394.3 & 2P(2,5) & 0.163$\pm$0.071 & 0.121$\pm$0.052 &
 ?6.4?$\pm$2.8? & ?2.25$\pm$0.09  \\ 
\hline
19& 399.8 & 2P(1,4) & 0.163$\pm$0.011 & 0.134$\pm$0.009 &
 ?7.2?$\pm$0.5? & ?2.54$\pm$0.24  \\ 
\cline{1-3} 
20& 405.9 & 2P(0,3) & (0.247$\pm$0.043)$^b$  & (0.205$\pm$0.035)$^b$  
 & (10.9?$\pm$1.9?)$^b$  &  \\ \hline 
\multicolumn{3}{|c|}{Sum(measured)}  & 3.40?$\pm$0.13? &  & &  \\ \hline
\multicolumn{3}{|c|}{Sum(corrected)} & 3.73?$\pm$0.15? &  & &  \\ \hline
\end{tabular}
\end{center}
\label{ta_comp}
\end{table}

In Table \ref{ta_comp}, we summarize the present results for
electrons of average energies of 0.85 MeV at 1000 hPa.  Fitted
values of $\epsilon$ in the 4th column correspond to photons in
each filter band, assuming only one line in each band, and
assuming a filter transmission at the wavelength of the main
line. The exception is the 391 nm filter, where $\epsilon$ is
separated into two lines.  The values in parentheses with upper
suffix $^a$ in column 4 are not measured in this experiment but
are estimated as follows.  The average ratio of our measured
values in other wave bands to the corresponding values listed in
Bunner \cite{bun64} is determined, and this ratio is multiplied
by the yields in Bunner.  In the case of the filters with central
values of 380.9 nm and 400.9 nm, the transmission coefficients for
the 375.6 nm and 405.9 nm lines are low compared to the values used
and the photon yields corrected for 
transmission coefficients of these lines 
are listed in parentheses and denoted by
an upper suffix $^b$.  The sum of the measured and corrected
values are listed at the bottom of column 4.

Values of $\Phi^{\circ}$ calculated from Eq.(\ref{eq-p}) are
listed in the 5th column.  In order to apply the present
measurements to air shower fluorescence experiments, the
fluorescence yields per meter per electron must be known as a
function of air density and the temperature for each wavelength.
$A_i$ and $B_i$ from Eq.(\ref{ep-temp}) are listed in the 6th and
7th column.

The total photon yield between 300 nm and 406 nm given by Bunner is
about 14 \% smaller than the present corrected results.  His value
was obtained as weighted averages of three measurements by
Davidson and O'Neil (50 keV electrons) \cite{dav64}, Hartman
(electrons, cited from \cite{bun64}) and Bunner (alpha particles)
\cite{bun64}, and the accuracy of each of the three experiments
was not better than $\pm$30 \% \cite{bun64}.  Therefore the
disagreement may be within their measurement errors.

\subsection{Effect of the present results on the energy estimation of UHECR}
\label{subsec:energy_est}

\begin{figure}[thb]
\centerline{\includegraphics[height=10cm]{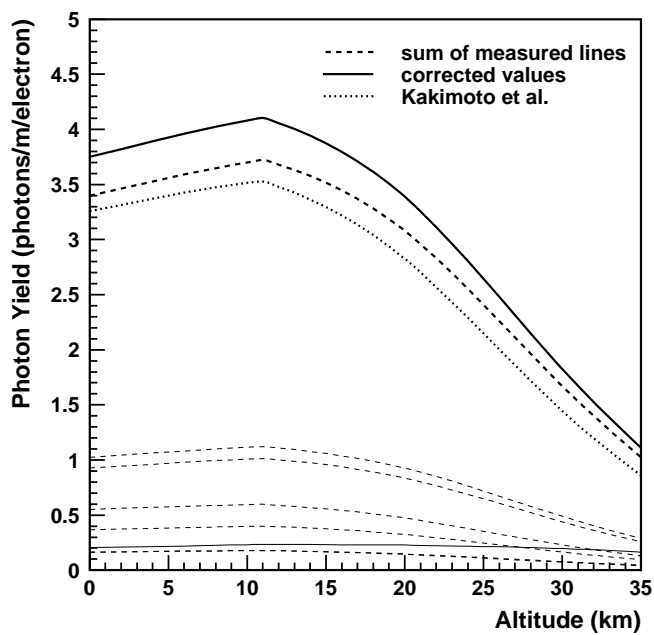}}
\caption{Photon yield between 300 and 406 nm for a 0.85 MeV electron
as a function of altitude.  The US Standard Atmosphere 1976
 \cite{rika02} is used 
for the atmospheric model.  The thin solid line is the altitude
dependence of the 391 nm line and the thin dashed lines are those of
other lines. The heavy dashed line is the sum of all measured lines. A heavy
solid line represents the corrected values for the unmeasured contributions.
The dotted heavy line is the photon yield between 300 nm and 400 nm
from Kakimoto et al., calculated for a 0.85 MeV electron using
their equation.}
\label{alt_photon}
\end{figure}

In Fig. \ref{alt_photon}, the photon yield between 300 nm and
406 nm for a 0.85 MeV electron is shown as a function of altitude.
The US Standard Atmosphere 1976 \cite{rika02} is assumed.  The
thin solid line is the altitude dependence of the 391 nm line and
the thin dashed lines are those of the other lines listed in
Table \ref{ta_comp}.  The heavy dashed line is the sum of all
measured lines.  It should be noted that the altitude dependence
is almost the same for all 2P lines.  The yield from the 391 nm line
is, however, almost independent of altitude, but its proportion
increases with altitude.  The heavy solid line is the corrected one
described in the previous section.  As done by Kakimoto et al.,
the total photon yields between 300 nm and 406 nm can be
approximated as a superposition of two sets of terms, for the 2P lines
and the 1N line, as follows:

\begin{equation}
\epsilon=\frac{(\frac{\d E}{\d x})}
                {(\frac{\d E}{\d x})_{0.85\mathrm{MeV}}}
          \times \rho \left(
      \frac{A_1}{1+\rho B_1 \sqrt{T}} + \frac{A_2}{1+\rho B_2 \sqrt{T}}
            \right) \ ,
\label{AandB}
\end{equation}
where $\rho$ is the air density in kg m$^{-3}$ and $T$ is in
Kelvin (K). Our values are $A_1=147.4\pm4.3 $ m$^2$kg$^{-1}$,
$A_2=69.8\pm12.2$ m$^2$kg$^{-1}$, $B_1=2.40\pm0.18$
m$^3$kg$^{-1}$K$^{-1/2}$, $B_2=20.1\pm6.9$
m$^3$kg$^{-1}$K$^{-1/2}$.  Here the energy loss
$\frac{\d E}{\d x}$ is normalized at 0.85 MeV.  Though
the parameters $A_1$, $A_2$, $B_1$ and $B_2$ from Kakimoto et
al. \cite{kak96} are different from the present results, their
altitude dependence is quite similar to the present experiment,
as shown by the heavy dotted line in the figure. But the absolute
values of the yield are somewhat different.

\begin{figure}[thb]
\centerline{\includegraphics[height=10cm]{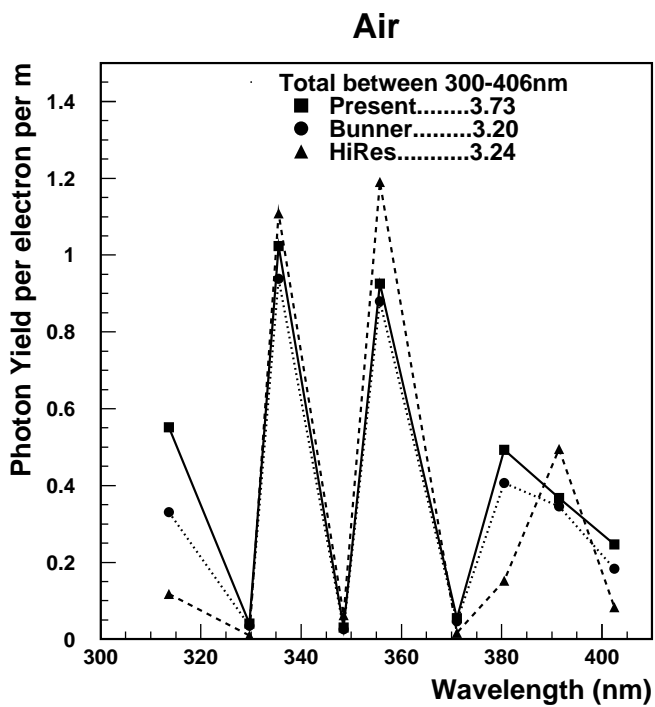}}
\caption{Comparison of $\epsilon$ from the present experiment
(solid line) with those from Bunner (dotted line) and those used
in the HiRes analysis (dashed line).}
\label{comp_hires}
\end{figure}

Fig. \ref{comp_hires} shows a comparison of $\epsilon$ from the
present results between 300 nm and 406 nm, with those from Bunner
and those values used by the HiRes experiment \cite{cao01}.
For an electron in air,  ($\frac{\d E}{\d x})_{0.85\mathrm{MeV}}$=0.1677
MeV/kg$\cdot$m$^{-2}$ and 
 ($\frac{\d E}{\d x})_{1.4\mathrm{MeV}}$=0.1659 
MeV/kg$\cdot$m$^{-2}$ corresponding to ours and
the HiRes, respectively. Therefore we may directly 
compare the present values of yields and those used by them.
Though the values used in the HiRes experiment are larger than
the present ones in the three main bands (337 nm, 356 nm, 391 nm),
the HiRes total number of photons between 300 and 406 nm is about
13 \% lower than ours as seen in Fig. \ref{comp_hires}.  
Though the systematic uncertainty of the present experiment
is 14\%, the main uncertainties are due to QE and CE and 
these measurement procedure by the factory is common to the present 
and Kakimoto et al..
Since the yields used in HiRes code is based on Kakimoto et al. 
\cite{cao01}, 
 the primary energy of UHECR may be estimated larger when
we use the HiRes photon yields than the present results.  
In order to know how much the
energy is largely estimated, it is necessary to take into account the
wavelength dependence of the attenuation of photons in the air
(Rayleigh and Mie scattering), the optical filter used, the quantum
efficiency of the PMT's and other factors which depend on the
wavelength.

Considering only Rayleigh scattering and the optical filter
transmission coefficient for the HiRes experiment, we have
estimated a change in energy of HiRes events if we use the
present results instead of their values.  The HiRes result may be
largely estimated by about 10 \% for showers of a distance around 10 km
and this value increases as the distance increases.  We must
evaluate the factor in more detail, by taking into account the
seasonal and daily variation of the density, temperature and the
Mie scattering at the experimental site, and the reconstructed
geometry of each shower track.

\section{Conclusion}
\label{sec:conclusion}
Photon yields have been measured in six wave bands as a function
of pressure, for nitrogen and dry air excited by electrons of
an average energy of 0.85 MeV.  The pressure dependencies of
fluorescence decay time were also measured.  The results are
summarized as follows:

 \begin{enumerate}
  \item Nitrogen results: The fluorescence efficiencies at 800 hPa
      are in good agreement with the results by Kakimoto et al.
      \cite{kak96,ueno96} at 337 nm and 358 nm, but 
      are smaller by a factor of 2.8 compared with the values of Davidson and
      O'Neil \cite{dav64} at 337, 358 and 380 nm. On the other hand,
      at 391 nm the present result agrees with theirs. 
  \item Air results : 
        The fluorescence efficiencies at 800 hPa are in good
        agreement with the previous experiments
        \cite{kak96,ueno96,dav64,har63}, if we take into account
        the backgrounds which were not subtracted in Kakimoto et
        al.  However, $\Phi^{\circ}$ for 391 nm is smaller, by a
        factor 3.5, than the measurement of Hirsh et
        al. \cite{hir70}, provided we use the best
        fit value of $p'$ from our experiment.

  \item The photon yield between 300 nm and 406 nm at 1000 hPa and
   20 $^\circ$C is 3.73$\pm$0.15 per meter for an electron of 0.85
   MeV.  The systematic error is 14 \%, with the main
   contribution relating to the collection efficiency of the PMT.
   The photon yield is proportional to
   $\frac{\d E}{\d x}$, and its density and temperature
   dependence is given by Eq.(\ref{AandB}).

  \item The photon yield between 300 nm and 406 nm used by the HiRes
   experiment is about 13 \% smaller than determined by the present
   experiment.  If we take into account the wavelength
   dependence of Rayleigh scattering and the transmission coefficient
   of the HiRes filter, the primary energies of UHECRs based on their
   photon yield is estimated larger than those based on the
   present experiment. The numerical factor
   depends on the distance between the shower trajectory and the
   experimental site.  

\end{enumerate}

\begin{ack}
We acknowledge M.Teshima of ICRR, University of Tokyo, for
helping us to prepare the experimental equipment and J.Hayami of
Fukui University of Technology for his kind support for this
experiment.  We are also grateful to B.Dawson of the University of
Adelaide for his improvement of the manuscript and his kind advice
and to the anonymous referee for the valuable suggestions and
comments.
This work is supported in part by the grant-in-aid for scientific
research No.12440068 from JSPS (Japan Society for the Promotion
of Science).
\end{ack}

\end{document}